\documentclass[useAMS,usegraphicx]{mn2e}
\usepackage{aas_macros}
\usepackage{rotating}

\title[A New Approach to Multiwavelength Associations]{A New Approach to Multiwavelength Associations of Astronomical Sources}
\author[I.G.Roseboom]{Isaac G. Roseboom$^{1}$\thanks{E-mail:
i.g.roseboom@sussex.ac.uk}, Seb Oliver$^{1}$, David Parkinson$^{1}$, Mattia Vaccari$^{2}$\\
$^{1}$Astronomy Centre, Dept of Physics and Astronomy, University of Sussex, Falmer, East Sussex, BN1 9QH\\
$^{2}$Department of Astronomy, University of Padova, Padova, Italy, I-35122\\
}
\begin{document}

\date{\today}

\pagerange{\pageref{firstpage}--\pageref{lastpage}} \pubyear{2008}

\maketitle

\label{firstpage}

\begin{abstract}
One of the biggest problems faced by current and next-generation astronomical surveys is trying to produce large numbers of accurate cross identifications across a range of wavelength regimes with varying data quality and positional uncertainty. Until recently simple spatial ``nearest neighbour'' associations have been sufficient for most applications. However as advances in instrumentation allow more sensitive images to be made the rapid increase in the source density has meant that source confusion across multiple wavelengths is a serious problem. The field of far-IR and sub-mm astronomy has been particularly hampered by such problems. The poor angular resolution of current sub-mm and far-IR instruments is such that in a lot of cases there are multiple plausible counterparts for each source at other wavelengths. Here we present a new automated method of producing associations between sources at different wavelengths using a combination of spatial and SED information set in the Bayesian framework presented by Budav\'{a}ri \& Szalay (2008). Testing of the technique is performed on both simulated catalogues of sources from GaLICS and real data from multi-wavelength observations of the SXDF. It is found that a single figure of merit, the Bayes factor, can be effectively used to describe the confidence in the match. Further applications of this technique to future Herschel datasets are discussed.

\end{abstract}


\section{Introduction}
Advances in astronomical instrumentation have simultaneously opened up new wavelength regimes while allowing deeper imaging capabilities in old ones. While this has allowed great advances to be made to our knowledge of the high redshift Universe, it has greatly increased the difficulty in producing accurate cross identifications between multi-wavelength datasets. The underlying causes for this are many; pushing to deeper flux sensitivities naturally results in a higher source density, while fundamental limitations to the angular resolution of imaging across all wavelengths means that more of these sources will become confused. This is particularly problematic when trying to make associations between deep optical/near-IR datasets and equivalently deep datasets at other wavelengths such as UV, far-IR/sub-mm or X-ray where the angular resolution is typically on the order of several to tens of arcseconds.

One area in which this has been a major stumbling block is the exploitation of the first observations in the sub-mm by ground based facilities around 850$\mu$m. Because of the strong negative k-correction at such wavelengths the brightest sources at 850$\mu$m are in fact high redshift ($z>2$) starburst galaxies (e.g. \nocite{Chapman2005}{Chapman} {et~al.} 2005) and hence will be optically quite faint. However current single dish sub-mm facilities, such as JCMT, APEX or even the IRAM 30m telescope, have apertures only in the 10s of metres, which results in typical 1$\sigma$ positional uncertainties of $\sim 4-6''$ for instruments such as SCUBA at 850$\mu$m (e.g. \nocite{Ivison2005}{Ivison} {et~al.} 2005). Herein lies the major difficulty in identifying counterparts to sub-mm sources, the density of sources at their predicted flux density in optical and near-to-mid IR bands is very high.

Ideally follow-up interferometric observations in the sub-mm with facilities such as the IRAM Plateau de Bure Interferometer (PdBI), the Submillimeter Array (SMA) and, in the near future, ALMA, could be used to reduce the positional uncertainty of sub-mm sources to match that of accompanying optical/near-IR data. However, given the small field of view and bandwidth constraints of such facilities, this is very observationally expensive and not a feasible option for the large number of sub-mm sources upcoming projects such as Scuba-2 and {\it Herschel} will produce.

Typically the approach to finding accurate positions for sub-mm sources has been to utilise deep interferometric radio observations, utilising the strong correlation between 1.4 GHz continuum flux and sub-mm flux \nocite{Ivison1998,Ivison2000,Smail2000,Ivison2002}({Ivison} {et~al.} 1998, 2000; {Smail} {et~al.} 2000; {Ivison} {et~al.} 2002, among many others). However this is also observationally expensive, the ratio between the sub-mm flux and the 1.4 GHz flux is expected to be around 100 at $z=2$. Given that the brightest sub-mm sources have a flux around 10mJy at 850$\mu$m, radio follow-up has to be deeper than at least $\sim 100\mu$Jy. Even for upcoming state of the art facilities this is a difficult proposition; E-VLA it will take 10-20hrs/sq. deg. to survey these sorts of depths. Thus it is clear that large areas of interferometric radio data of adequate depth will not be readily available for some time.

On top of this deep radio data counterparts are often not found, indeed there is some evidence that the most luminous sub-mm galaxies are also radio-dim \nocite{Younger2007,Ivison2002}({Younger} {et~al.} 2007; {Ivison} {et~al.} 2002).

Identification of counterparts in mid-IR observations from {\it Spitzer} has been attempted by several authors. \nocite{Ivison2004}Ivison et al. (2004) and \nocite{Egami2004}Egami et al. (2004) where amongst the first to try and utilise deep {\it Spitzer}, in particular MIPS 24$\mu$m, imaging to find counterparts for sub-mm sources. Ivison et al. identified reliable \footnote{Here and throughout we define reliable to mean $<5$\% chance of a spurious association} 24$\mu$m counterparts for 8 out of 9 $>3\sigma$MAMBO 1200$\mu$m sources. Egami et al. similarly found 24$\mu$m identifications for 7 out of 10 $>3\sigma$ SCUBA sources, with the remaining three lacking both radio and 24$\mu$m counterparts. Further work paints a similar picture, with 24$\mu$m counterparts for most sources, but generally only those which are also quite strong in the radio. Using deep 8$\mu$m IRAC data \nocite{Ashby2006}{Ashby} {et~al.} (2006) were able to identify reliable counterparts for 17 SCUBA detected SMGs in the CUDSS 14 hour field. Of these only 5 had previous 1.4 GHz radio counterparts from relatively shallow imaging ($\sim$60$\mu$Jy), highlighting the usefulness of shorter wavelength identifications in the absence of good radio data. \nocite{Ivison2007}{Ivison} {et~al.} (2007) found statistically significant 24$\mu$m counterparts for 53 out of 120 SCUBA sources from the SHADES survey of SXDF and Lockman Hole, however only 11 of these were previously undetected in deep radio data. 

In both the mid-IR and radio identifications the ``goodness'' of an association is determined using the $p$ statistic \nocite{Lilly1999,Ivison2002}({Lilly} {et~al.} 1999; {Ivison} {et~al.} 2002), which is defined as the probability that a radio or mid-IR source of a particular flux density could be found {\it by chance} at particular distance from the sub-mm source. In this way most catalogues of multi-wavelength associations to sub-mm sources are constructed by taking those with radio/mid-IR associations that have $p<0.05$, i.e. they have a less than a 5\% chance of being spurious. This approach is very useful for finding radio associations, as the density of $\mu$Jy 1.4 GHz radio sources is still quite low compared to the typical positional uncertainty in the sub-mm. However in the mid-IR similarly faint 8 or 24 $\mu$m sources are numerous enough that only strong or very nearby sources have sufficiently small $p$ statistics to be considered as confident associations.  

One thing that is clear is that with the advent of new wide areas surveys in the sub-mm using facilities such as BLAST \nocite{Pascale2008}({Pascale} {et~al.} 2008), SCUBA-2 and {\it Herschel} deep radio data will not be readily available for the significant numbers of SMGs that will be detected. Thus an alternative approach which is able to utilise the multi-band data at hand are needed.
  
Here we present a new technique to narrow down the number of potential matches using the Bayesian statistical framework found in Budav\'{a}ri \& Szalay (2008) \nocite{Budavari2008}. Importantly our technique considers both the SED and spatial information in determining which combination of multi-wavelength data is associated with a sub-mm/far-IR source. The formalism of this new technique is presented in Section \ref{sec:technique}. The real and simulated datasets utilized in testing this technique are described in Section \ref{sec:data}, with the results of these tests presented in Sections \ref{sec:scubsims} \& \ref{sec:shades}, respectively. Finally Section \ref{sec:discussion} discusses the benefits of this technique, while Section \ref{sec:herschel} demonstrates its applicability to upcoming {\it Herschel} datasets.

\section{Cross-identification technique}\label{sec:technique}
Our association technique is broken down into a two step process. First spatial matching is performed to find potential associations between a sub-mm source and the objects in the catalogues at other wavelengths. This process is performed as per the iterative technique presented in Budav\'{a}ri \& Szalay (2008; henceforth BS08). The BS08 approach relies on Bayesian hypothesis testing, where the hypothesis under consideration is that $n$ sources from $n$ catalogues at different wavelengths originate from the same astronomical object. This is compared to the alternative hypothesis that the $n$ sources come from $n$ different astronomical objects. The Bayes factor is essentially the ratio of the posterior and prior probabilities of these two scenerios. The full mathematical basis for this technique is summarised in Appendix \ref{app:bsmath}.

One major disadvantage of the BS08 approach is that it only considers the as an alternative hypothesis that the $n$ sources come from $n$ different physical objects, ignoring the likely scenario that some, but not all, of the sources from different catalogues are associated.  In sub-mm/far-IR astronomy we are typically trying to associate one set of sources with poor positional uncertainties (i.e. our sub-mm source catalogue) to sets of sources with very accurate positional information (i.e. optical or interferometric radio source catalogues). Thus forming reliable associations between the high resolution data is relatively easy and can be accomplished with simple nearest neighbour techniques. The real challenge is establishing the link between these associations and our poorly resolved sub-mm/far-IR sources. Given this we take a slightly different philosophical approach than BS08. Rather than consider that the $n$ sources come from $n$ distinct objects as an alternative hypothesis we consider the $n$ sources come from 2 distinct objects; the $n-1$ high-resolution sources from one astronomical object and the sub-mm/far-IR sources from another. While this does not affect the calculation of the Bayes factor for the spatial associations, it is simply the same as considering the inputs as two catalogues as opposed to $n$, it has a fundamental affect on the way we calculate the Bayes factor for the SED. Here we calculate the Bayes factor for the SED by comparing the hypothesis that an object has the measured flux (H) to the alternative hypothesis that it actually has a flux below the detection limit (K). Mathematically we calculate this via the Bayes factor;
\indent \[B_{HK}=\frac{\int{p(n|H)p(g|n,H)dn}}{\int{p(n|K)p(g|n,K)dn}}\]
where $n$ represents the parameterisation (T,$z$,$A_v$) of each template T, redshift $z$ and extinction $A_v$ considered. $p(g|n,M)$ is the probability of hypothesis $M$, $p(n|M)$ is the prior probability on $M$ and the integrals run over the range of models, redshifts, and dust extinction. 


Here we assume that the likelihood of a SED being ``correct'' is given by the $\chi^2$ distribution via 
\indent \[p(g|n,M) \propto \exp\{-\sum_{i=1}^{N}{\frac{[g_i-bf_i(T,z,A_v)]^2}{2\sigma_{g_i}^2}}\}\]

\noindent where $g_i$ is the observed flux in passband $i$, $f_i$ the model flux in $i$ given (T,$z$,$A_v$), b the normalisation factor and $\sigma_{g_i}$ the error on $g_i$. $N$ is the number of observed bands.

In instances where candidate matches are undetected in one or more band the flux limits are introduced as a proxy for the measured flux. If the model flux for passband $i$ is below the flux limit it is not considered in the sum (i.e. $g_i-bf_i(T,z,A_v)=0$). However if the model flux for passband $i$ is greater than the flux limit the flux limit is used, with the error assumed to be the typical error for sources near the limit. 


 The SED fitting we perform follows a similar prescription to the photometric redshift estimation technique described in Rowan-Robinson et al. (2008; Henceforth RR08)\nocite{Rowan-Robinson2008}, however with several key differences. The subtle difference between normal photometric redshift estimation and the technique used here is that we are performing Bayesian hypothesis testing, not parameter estimation. Hence the actual best fit parameters (i.e. redshift, template, A$_v$) are not the goal here, the aim is to statistically test whether or not a particular combination of sources from different catalogues is responsible for the flux detected. We utilise a subset of the templates from RR08 to fit our candidate matches. For the optical to near-IR only the 7 galaxy templates from RR08 are considered; 2 elliptical, 5 spiral types from Sa to Sdm. In the far-IR we consider a Arp220, M82 and Cirrus template, again taken from RR08. While more templates, and indeed combinations between templates, are typically required to produce good SED fits we find that giving the fitting process this additional freedom makes it easier to obtain ``good'' fits (i.e. low $\chi^2$) to clearly mismatched sources. 

Redshifts in the range 0-4 are considered, with a step size of 0.002 in Log$(1+z)$. Dust extinction in the range $0<A_v<1$ is also considered, with the form of the extinction as per Calzetti et al. (2000) \nocite{Calzetti2000}. 

The fitting process itself is performed via the least squares fitting of two components (optical + far-IR templates) to the observed fluxes via the use of a non-negative least squares fitting algorithm, in this case the Bounded Variable Least Squares (BVLS) algorithm \nocite{Stark1995}(Stark~P.B. 1995)).

As we wish to demonstrate the general applicability of the technique, a minimum level of priors is assumed. Of course the selection of the SED templates is in itself a very strong prior, however as the templates used in this work have been shown by RR08 to match a large fraction of the SWIRE galaxy population it is reasonable to believe that they represent a fair sampling of the underlying galaxy population. In addition to ensure that we are not assigning statistical significance to implausible solutions a luminosity prior is included in the same fashion as RR08, i.e. $-17-z>$M$_{B}>22.5-z$ for $z<2$ and $-19.5>$M$_{B}>-25$ for $z>2$.


In practice it is impractical to consider {\it every} combination of {\it every} source in the input catalogues. Thus at each step combinations of sources which are greater than some arbitrary search radius (typically 3-5 times the assumed positional error) can be excluded from the calculations. After a list of candidate spatial matches has been compiled using the BS08 formalism SED fitting is performed on each to try and find the best match to the source.


Thus the algorithm for determining the final association is as follows;

\begin{enumerate}
 \item calculate ln B$_{spatial}$ and ln B$_{sed}$ for each association within an arbitrary radius.
 \item Add together ln B$_{spatial}$ and ln B$_{sed}$ to give the final Bayesian evidence, ln B$_{tot}$.
 \item Find the largest value of ln B$_{tot}$ out of the potential matches. This is the final association.
\end{enumerate}

\section{Data} \label{sec:data}
To demonstrate the effectiveness of our technique we make use of both simulated data from GaLICS \nocite{Hatton2003}({Hatton} {et~al.} 2003) and real observations of the Subaru-XMM Deep Field (SXDF) from ground and space based facilities . 

The GaLICS simulations are well-suited to our purposes as they incorporate realistic clustering, realistic star formation histories and galaxy properties, and simulated SEDs which cover a wavelength range from UV through to sub-mm.

The real data focuses on the SHADES SCUBA observations in the Subaru XMM-newton Deep Field (SXDF). This dataset has the advantage of having accompanying ancillary data at optical, IR and radio wavelengths allowing high precision multi-wavelength associations to be made for a large number of the SCUBA sources \nocite{Ivison2007,Clements2008}({Ivison} {et~al.} 2007; {Clements} {et~al.} 2008). This allows us to test out optical to far-IR identifications against a ``truth'' list of radio identifications.

The SHADES survey performed 850$\mu$m observations with SCUBA on the James-Clark Maxwell Telescope (JCMT) of a 0.2 sq. deg pointing coincident with the Subaru XMM-newton Deep Field (SXDF). Details of the observations and resulting maps and catalogues can be found in \nocite{Coppin2006}{Coppin} {et~al.} (2006).

For the SHADES SCUBA associations we utilise data from a number of deep surveys in the SXDF region. 
In the optical we utilise public DR1 release of the Subaru XMM-newton Deep Field Survey \nocite{Furusawa2008}({Furusawa} {et~al.} 2008). SXDS observed 5 fields in a 'plus' shaped pattern centered on Right Ascension=$02^h18^m00^s$ and Declination=-$05^{\circ}00'00''$ with the SuprimeCam instrument on the Subaru telescope. The SHADES field is wholly contained within the single central SuprimeCam pointing. Observations were performed in 5 optical bands, B, V, R, $i'$, $z'$, with 3$\sigma$, 2'' aperture, AB mag depths of 27.5, 27.5, 27, 27, and 26 respectively.

In the near-IR we utilise data from the UKIDSS (UKIRT Infrared Deep Sky Survey; \nocite{Lawrence2007}{Lawrence} {et~al.} 2007) Ultra Deep Survey (UDS). UKIDSS uses the UKIRT Wide Field Camera (WFCAM; Casali et al, 2007)\nocite{Casali2007} and a photometric system described in Hewett et al (2006)\nocite{Hewett2006}. The pipeline processing and science archive are described in Irwin et al (2008, in prep) and Hambly et al (2008)\nocite{Hambly2008}. We utilise the DR3 release of the UDS dataset, which contains photometry in J,H and K to a 5$\sigma$ depth of 23.7, 23.5, and 23.7 AB mags respectively. The UDS field is coincident with the SXDF, covering the extent of the SHADES SCUBA observations. 

For the mid-far IR we utilise data from the SWIRE  survey. SWIRE contains imaging of the entire XMM-LSS field in both the IRAC and MIPS instruments on SWIRE. This results in a 5 band dataset, with flux measurements centred on 3.6, 4.5, 5.8, 8.0, 24.0 $\mu$m. While the MIPS 70$\mu$m and 160$\mu$m data is included in the analysis only a very small number of SHADES sources are found to have nearby 70 and/or 160$\mu$m sources in SWIRE and thus it is of little use in the vast majority of cases.

\section{Testing on simulated catalogues: GaLICS}\label{sec:scubsims}
As an initial test we try to recreate in simulated data the real scenario that will be considered later in this paper the matching of 850$\mu$m sources to deep optical and {\it Spitzer} IRAC \& MIPS data with no redshift information. The use of the simulations at this stage is vital as it offers the convienence of a perfect truth list to test against, i.e. we know the true underlying association for each object apriori, something which is never truly possible with real data.

We select 1 cone (Cone 1: 1 sq. deg.) of GaLICS simulations with photometry in 5 optical bands (B,V,R,i',z'), 3 Near-IR bands (J,H,K), the 4 {\it Spitzer} IRAC bands, and the MIPS 24 \& 70$\mu$m bands. The simulated data is then broken up into three catalogues, an optical-near IR catalogue, a ``SWIRE'' Spitzer IRAC+MIPS 24$\mu$m catalogue and a SCUBA 850$\mu$m catalogue. Flux limits are introduced to make these catalogues resemble those found in the SXDF. All objects with $B<27$, S$_{3.6}>10\mu$Jy and S$_{850}>$2 mJy are kept in the catalogues. Additional limits are placed on the fluxes in the ``SWIRE'' catalogue to take into account the varying sensitivity between the IRAC channels and MIPS 24 $\mu$m. Catalogued objects with flux values less than 40$\mu$Jy at IRAC 5.8$\mu$m \& 8.0$\mu$m,  or $<50\mu$Jy at 24$\mu$m are treated as undetected at these wavelength in the analysis. While these limits are somewhat lower than the real data in SXDF they better match the observed number density of objects in each catalogue. This mismatch is a result of a natural disparity between the number density of far-IR luminous sources in the GaLICS simulations compared to the real Universe.

These cuts result in catalogues of 253 SCUBA 850$\mu$m sources, 34932 Spitzer sources (10817 with 24$\mu$m), and 306842 optical+near-ir sources, respectively. 

As the flux limits for each catalogue are imposed in different wavelength regimes there is a natural disparity between the catalogues, this is reflected by the fact that not all of the sources in one catalogue have matches in the other two. Specifically, only 129 of the 253 mock SCUBA sources have corresponding entries in the {\it Spitzer} catalogue. Of these 117 are ``detected'' at 24 $\mu$m (i.e. model S$_{24\mu m}>50\mu$Jy).

The positions of objects in the three catalogues are independently scattered by Gaussian random errors, with the positional uncertainty in each case being: Optical 0.1'', Spitzer 0.2'' and SCUBA 850$\mu$m 3''. 

For our first test we try and find associations between our three catalogues requiring that a 24$\mu$m detection is present in the {\it Spitzer} catalogue (i.e. S$_{24\mu m}>50\mu$Jy). For comparison we also find the best 24$\mu$m association for each mock SCUBA source using the p-statistic \nocite{Downes1986}({Downes} {et~al.} 1986). Table \ref{tab:sims24um} summarises the our results in terms of completeness (total number of correct matches over all true associations) and reliability (number of correct matches over total made). Like the p-statistic our approach relies on a single statistic to determine the believability of an association; the Bayesian evidence $\ln B$. Three $\ln B$ selection thresholds are presented; none, $\ln B>5$, and $\ln B>2.2$. The $\ln\ B>5$ selection is consistent with ``strong evidence'' for a match according to the Jeffery's' scale \nocite{Jeffreys1961}Jeffreys (1961). A final selection ($\ln B>2.2$) which matches the number of associations found via the $p<0.05$ selection is also shown. This is to enable a fair and direct comparison between the two methods.

\begin{table}

\caption{Summary of completeness (C) and reliability (R) of matching between simulated optical, {\it Spitzer} and SCUBA band catalogues, where we require that the association has a measured flux at 24$\mu$m.}
\label{tab:sims24um}
\begin{tabular}{l|l|l|l|l|}

\hline
 & Total & Correct & C & R\\
\hline
$p_{24}<0.05$ & 106 & 92 & 79\% & 86\%\\
Bayesian Matching (no cut) & 181 & 115 & 98\% & 64\%\\
Bayesian Matching ($\ln B>5$) & 89 & 85 & 73\% & 96\%\\
Bayesian Matching ($\ln B>2.2$) & 106 & 96 & 83\% & 90\%\\

\end{tabular}
\end{table}

\begin{figure}

\includegraphics[angle=270,scale=0.6]{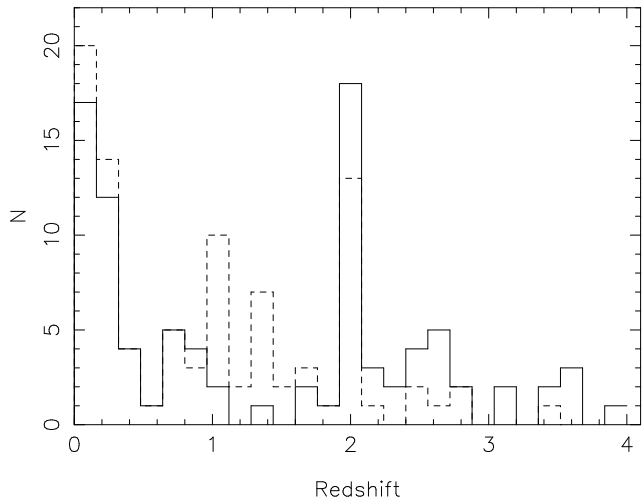}
\caption{Redshift distribution of correctly recovered associations using both our Bayesian approach (solid line) and the p-statistic (dot-dashed line). The Bayesian approach appears to recover more high-$z$ ($z>1.5$) associations, while the p-statistic is more effective at low to intermediate redshift. This can be attributed to the different philosophy of each approach, as explained in the text.}
\label{fig:simsNz}
\end{figure}

From Table \ref{tab:sims24um} it can be seen that the Bayesian analysis performs similarly to the p-statistic in correctly associating sub-mm to shorter wavelength counterparts, with both achieving a completeness of around 80\% and a reliability of $\sim$90\%. One advantage of the Bayesian approach over the p-statistic is its ability to make correct identifications at the highest redshifts. This can be seen in Figure \ref{fig:simsNz} which shows the redshift distribution for the correct Bayesian and p-statistic matches. The Bayesian approach correctly recovers 38/42 associations at $z>1.5$ while the p-statistic only recovers 20/42. The reason for this is clear, the 24$\mu$m flux of these sources drops dramatically as a function of redshift, whereas the 850$\mu$m flux, as a result of the negative $k$ correction, stays the relatively constant. Thus the 24$\mu$m associations for high-$z$ SCUBA sources will always be very faint, and hence have a high number density. As the p-statistic is based on the number density of sources at or above a given flux level it will often determine that faint, higher-$z$, counterparts have a high chance of being spurious. 

However Figure \ref{fig:simsNz} does show one failure of the Bayesian approach, which is a decrement of correct associations around $z=1-1.4$. This can be attributed to the fact that the GaLICS simulated SEDs do contain as strong absorption in the silicate feature at rest frame $9.7\mu$m as found in the RR08 templates. As this enters the 24$\mu$m band at $z\sim1$ the ``observed'' 24$\mu$m flux is always found to be much greater than would be expected from the RR08 templates. This results in poor fits to the available templates, in turn causing associations whose true redshift is in this range to be given low evidence.  

While associations with 24$\mu$m sources are a reasonably reliable ``gateway'' to associations at shorter wavelengths unfortunately not all sub-mm sources will have detections at 24$\mu$m. Thus we would like to be able to find reliable associations for these sources on the basis of {\it Spitzer} IRAC and optical/near-IR data alone. To determine if our approach can successfully produce associations in the absence of 24$\mu$m detections we repeat the analysis above, but this time allow sources from the {\it Spitzer} catalogue without 24$\mu$m to be considered as potential counterparts. The completeness and reliability of these associations is given in Table \ref{tab:simsall}

\begin{table}
\caption{Summary of completeness (C) and reliability (R) of matching between simulated optical, {\it Spitzer} and SCUBA band catalogues, where we do not require that the association has a measured flux at 24$\mu$m.}
\label{tab:simsall}
\begin{tabular}{l|l|l|l|l}
\hline
Base priors & Total & Correct & C & R\\
\hline
$p_{24}$ or $p_{3.6}<0.05$& 109 & 93 & 72\% & 85\%\\
Bayesian Matching (no cut) & 225 & 115 & 89\% & 51\%\\
Bayesian Matching ($\ln B>5$)& 130 & 93 & 72\% & 72\%\\
Bayesian Matching ($\ln B>8$)&109 & 85 & 66\% & 78\%\\
\hline
\multicolumn{5}{l}{Redshift and extra M$_{B}$ prior}\\
\hline
Bayesian Matching (no cut) & 225 & 117 & 91\% & 52\%\\
Bayesian Matching ($ln\ B>5$)& 110 & 94 & 73\% & 85\%\\
Bayesian Matching ($ln\ B>5.2$)&109 & 93 & 72\% & 85\%\\
\end{tabular}
\end{table}

By allowing associations between sub-mm and IRAC only (i.e. 3.6$\mu$m) sources the reliability of our associations drops considerably. For comparison the p-statistic is also computed for the 3.6$\mu$m sources, with the final p-statistic determined association the better of the 3.6$\mu$m or 24$\mu$m associations, where available.

Under these circumstances our approach results in considerably worse completeness and reliability than the p-statistic for the same number of objects selected. This is not unexpected, the number of 850$\mu$m sources in the simulations without accompanying 24$\mu$m detections is very small (12/129), while the number of {\it Spitzer} catalogue objects without 24$\mu$m is much higher (24115/34932). 

Encouragingly we correctly identify all 12 850$\mu$m sources without 24$\mu$m counterparts with strong evidence ($\ln B>9$), while the p-statistic only recovers one of these. 

\begin{figure}

\includegraphics[angle=270,scale=0.6]{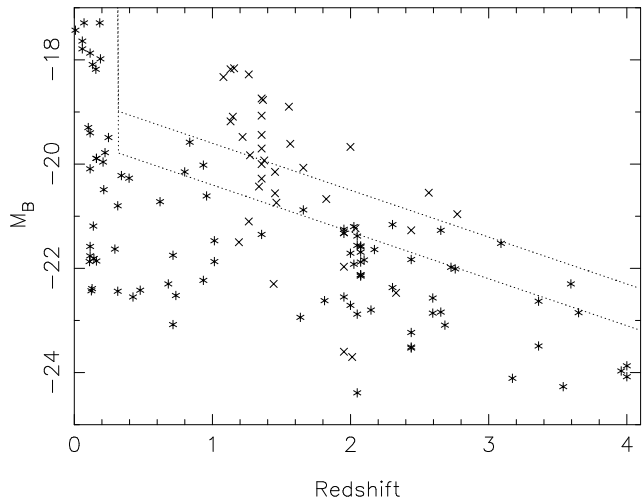}
\caption{M$_B$ vs. redshift for correct (astericks) and incorrect (crosses) associations with strong evidence ($\ln B >5$). The vast majority of incorrect associations are found in the redshift range $1.2<z<1.6$ with low M$_B$. The parameter space between the dotted lines represents the region which we attempt to disfavour in the SED fitting via the introduction of the prior discussed in the text. }
\label{fig:MBzsims}
\end{figure}

Investigating the properties of the incorrect associations made by our technique immediately reveals the reason for such poor performance in this scenario. Figure \ref{fig:MBzsims} shows the M$_{B}$ vs. redshift for the $\ln B >5$ associations. The vast majority of the mismatches are located in the redshift range $z=1.2-1.6$, with systematically lower optical luminosities. This failure of our approach can be attributed the erroneous treatment of the $9.7\mu$m silicate absorption feature in the model SEDs. While in the case of the 24$\mu$m only associations this hampered our ability to make associations with strong evidence, here 3.6$\mu$m associations with strong evidence are able to be made made as the predicted 24$\mu$m flux from the templates is much lower, and hence much closer to (or below) the S$_{24\mu m}>50\mu$Jy flux limit. Knowing this we can introduce a prior based on the M$_{B}$--$z$ evolution such that these low-luminosity, intermediate redshift solutions are strongly disfavoured. The prior introduced is defined as $p=1-(M_B+19.5+0.9*z)$ for $-0.9*z-19.5<M_B<-0.9*z-18.7$ \& $z>0.3$, $p=1$ for $M_B<-0.9*z-19.5$ or $z<0.3$ and $p=0$ for $M_B>-0.9*z-18.7$ \& $z>0.3$. In a further effort to recover the true associations we also introduce the true redshift distribution of the 850$\mu$m sources as a prior, where the redshift distribution of sources is modelled as a Gaussian with mean $z=2.24$ and $\sigma=0.945$. Again this prior is only invoked at $z>0.3$ so as not to affect our ability to recover the small number of low luminosity 850$\mu$m sources at low redshift. The result of repeating the analysis with the introduction of these priors is given in Table \ref{tab:simsall}.

As with the 24$\mu$m only associations we succeed in doing as well as a p-statistic based analysis, with the major benefit being that larger number of high-$z$ associations made. In this scenario 46/53 $z>1.5$ 850$\mu$m sources are correctly identified with strong evidence, while only 24/53 are recovered by the p-statistic.

\section{Testing on SHADES Scuba sources}\label{sec:shades}
To test the effectiveness of our association technique on real data we try to reproduce the optical--mid-IR associations of SHADES SCUBA sources with confident radio IDs in SXDF compiled by Ivison et al. (2007; hereafter I07) and subsequently analysed to produce photometric redshifts and SED fits by Clements et al. (2008; hereafter C08)\nocite{Ivison2007,Clements2008}. Here we propose a simple test for the application of our association algorithm; produce associations with only the optical and infrared datasets and see if these coincide with the 33 SMGs with confident radio associations presented in I07 \& C08. Table \ref{tab:scubaresults} compares the associations and best-fit photometric redshifts from I07 \& C08 to those determined using our Bayesian approach. 

Several important caveats must be introduced before the matching can be performed. Firstly only SWIRE catalogue objects with detections in 2 or more bands are considered. This restriction is imposed as a faint single band SWIRE detections is highly likely to be spurious, and also easily fit by any far-IR SED. As we are now dealing with real data the redshift and the extra M$_{b}$ priors introduced in the previous section should now longer be necessary as these were introduced to overcome deficiencies in the model SEDs. Thus the only priors included are the M$_{B}$ prior from RR08 to ensure plausible luminosities, and the implicit prior that the SEDs are well approximated by our limited range of templates. In addition as we expect the SED fitting to be more successful on the real data we consider all matches, not just those with 24$\mu$m IDs. While any associations made without the benefit of a 24$\mu$m detection should be treated with great care, it will be very interesting to see if our technique can be successful in making reliable associations with IRAC \& near-ir/optical data alone.

To account for the limited range of templates and systematic differences between these and the data, a minimum error is enforced on all of the flux densities. For the optical and UKIDSS near-IR data a minimum error of 0.05 mags is enforced. For the IRAC data a minimum error of 5\% error is used, while for the MIPS \& SCUBA data a minimum error of 10\% is used. In addition only optical/IR sources within a 15'' radius of the SCUBA source are considered. While in principle this technique should consider sources at all separations, for practical reasons, i.e. the limitation of computer power, it is necessary to impose a maximum search radius. This radius was chosen as it is well matched to the beam of the SCUBA instrument at 850$\mu$m and safely encompasses all the associations presented in C08.

\begin{table*}

\caption{Comparison of Scuba radio associations presented by Ivison et al. (2007) \& Clements et al. (2008:C08) to those determined here. Column 2 details whether or not the association from I07 is reproduced; Y for yes, N for no and an incorrect association has been made. Where the incorrect, or an alternative, association has been made the values for the C08 association(s) are shown (when possible). Seperation quoted is distance from best-guess position to 850$\mu$m source in arcseconds.}
\label{tab:scubaresults}
\begin{tabular}{l|l|l|l|l|l|l|l|l|l}
\hline
SHADES-SXDF ID & \multicolumn{2}{c}{Position} & Agree? (Y/N) & C08 z$_{phot}$ & z$_{phot}$ & $\ln B_{tot}$ & $\ln B_{sed}$ & $\chi^2$ & Separation (``)\\
& RA (deg) & Dec (deg) & & & & & & & \\
\hline
1 & 34.3733 & -4.99366 &  N & 1.44 & 1.18 & -34.0 & -12.6 & 51.8 & 14.1\\
1 (C08)  & 34.3776 &-4.99354 &  &      & 0.8 &  &  &  & 1.30\\
2  &34.5149 & -4.92432  & Y & 2.39 & 3.19 & 52.3 & 38.2 & 97.1 & 1.0 \\
3 & 34.4224 &-4.94324 & N & 0.41 & 1.2 & -5.1 & -10.4 &40.5 & 13.8\\
3 (C08)  & 34.4256  & -4.94124 &  &      & 0.49 & -76.2 & -90.3 & 38.7 & 0.26\\
4 & 34.4112 &-5.06096 &Y & 2.22 & 2.29 & 15.2 & 1.3 & 26.4 & 2.3\\
5 & 34.5119 &-5.00857 &Y & 1.67 & 3.0 & 89.9 & 76.0 & 211.8 & 2.0\\
6 (C08:603) & 34.3742 &-5.05531  &Y & 1.1 & 2.57 & 33.8 & 22.3 & 50.9 & 7.7\\
7 & 34.4120 & -5.09112   &Y & 2.31 & 2.87 & 20.4 & 7.1 & 91.6 & 4.4\\
8 &  34.4340 &-4.93029 &Y & 2.53 & 1.7 & 34.7 & 22.9 & 254.5 & 7.0\\ 
10 & 34.6053 & -4.93285 &N & 2.08 & 1.16 & 14.7 & 0.7 & 26.2 & 1.16\\
10 (C08)   &  34.6062 &  -4.93303 & &  & 1.49 & & & & 3.95 \\
11 & 34.3548 & -4.99300 &Y & 2.3 & 0.5 & 9.5 & -4.3 & 12.7 & 2.6\\
12 & 34.4943 &-5.08449  &N & 3.07 & 1.0 & 7.1 & -1.7 & 35.0  & 11.15  \\ 
12 (C08)   &  34.4971 &  -5.08447& & & 0.5 & & & & 1.1\\
14$^{1}$ &34.5780 & -5.04706 & Y & 2.24 & 2.45 & 8.1 & -2.1 & 57.9 & 9.5\\
19 &  34.6161 & -4.97689 &Y & 2.06 & 2.34 & 14.4 & 1.4 & 22.7& 4.9\\
21$^{2}$ &34.4271 &-5.07359 &Y & 0.044 & 0.13 & 704.8 & 691.8 & 95.3 & 5.1\\
23 & 34.4268& -5.09604 &Y & 2.22 & 3.36 & 8.6 & -5.4 & 27.0 & 1.5 \\
24$^{1}$ & 34.3947& -5.07502 &Y & 1.01 & 1.08 & 1.9 & -9.5 & 75.3& 8.0\\
27 (C08:2701) & 34.5330 &-5.02931 &Y$^{3}$& 2.0 & 3.2 & 34.6 & 20.9 & 28.1 & 3.1\\
27 (C08:2703)   &   34.5333 &  -5.03169&       &      & 1.3& 5.7 & -6.8 & 212.3 & 5.93\\
27 (C08:2702)   &   34.534  &  -5.02791 &      &      & 1.4 & 4.0& -6.5 & 164.5 & 9.1 \\
28 &34.5284 & -4.98821 &Y$^{4}$ & 1.16 & 1.2 & 21.4 & 8.1 & 595.2 & 3.9\\ 
28 (C08)   & 34.5287&   -4.98692 &       &       & 0.98 & -8.0 & -21.6 & 317.6 & 3.48\\
29 (C08:2902)& 34.5672 &-4.91921 &N & 1.0 & 1.63 & -593 & -605 & 732.9 & 5.7 \\
29 (C08:29+) & 34.5672 & -4.91915 & & & & & 6.1\\
30 &34.4167& -5.02101 &Y & 1.13 & 3.1 & 52.8 & 39.7 & 43.6 & 4.58\\
31 (C08:3101) &  34.3994 & -4.93208  &Y$^{4}$ & 2.21 & 0.81 & 7.1 & -4.6 & 43.6 & 7.0\\
31 (C08:3102) & 34.4025  & -4.93201     &        &      & 1.6 & -4.9 & -17.8 & 40.9 & 5.1\\
35 &    34.5020& -4.88686   &N & 1.23 & 1.0 & 3.6 & -8.6 & 1271.0 & 6.3\\
35(C08) & 34.5035 &  -4.88503     &  &      & 1.17 & -90.6 & -103.4 & 165.6 & 5.1\\
37 & 34.3524 & -4.97812 &Y & 1.34 & 0.2 & 8.5 & -5.3 & 317.7 & 2.4\\
47 (C08:4701) & 34.3932 & -4.98258  &Y$^{4}$ & 1.39 & 1.04 & 8.3 & -3.5 & 395.5 & 7.1\\
47 (C08:4702)   & 34.3904 &  -4.98268 &        &      & 1.32 & -274.0 & -287.7 & 393.4 & 3.0\\
47 (C08:4703)   &  34.3933 &  -4.98327   &       &      & 1.3 & -16.3 & -27.5 & 54.5 & 7.9\\
52 (C08:5202)$^{5}$ & 34.5206 &-5.08367   &Y$^{3}$ & 2.98 & 1.4 & 9.1 & -2.6 & 693.0 & 7.6\\
52 (C08:5201)   &34.5213 &  -5.08155   &        &      & 0.5 & 3.6 & -10.0 & 177.7 & 3.2\\
69$^{1}$ &  34.4660& -5.04958  &N & 1.31 &  0.7   & 5.3     &   -4.3    & 262.7 & 10.1  \\
69 (C08)&  34.4627 &  -5.05077  &      & 1.05 & -7.8 & -14.6 & 96.3 & 13.0\\
71 & 34.5887& -4.98366  &Y & 2.24 & 3.5 & 10.3 & -3.6 & 86.9 & 2.1\\
74$^{1}$ & 34.4927 & -4.90575   &N & 0.75 & 3.1 & 6.1 & -2.2 & 27.4 & 10.9\\
74 (C08)         &  34.4952 &   -4.90863  &  &      & 0.73 & -1.3 & -14.9 & 20.7 & 2.7\\ 
76 & 34.4846 & -5.10692  &Y & 1.07 & 1.0 & -55.2 & -66.3 & 88.2 & 8.5\\
77 (C08:7701) &  34.4007 & -5.07598  &Y$^{3}$ & 0.98 & 0.99 &-4.1  & -17.5 & 110.3 & 4.2\\
77 (C08:7702)   &34.3997  & -5.07386   &        &      & 0.85 & -5.2 & -15.2 & 49.8 & 9.8\\
88$^{1}$ &  34.5029 & -5.08148 &N & 1.34 & 1.3 & 6.1 & -6.2 & 26.9 & 6.5\\
88 (C08)        & 34.5067 &  -5.0786  &   &     & 0.67 & -4.9 & -14.1 & 22.5 & 10.7\\
96 & 34.5008& -5.03786 &Y & 1.78 & 1.63 & 9.1 & -4.1 & 27.9& 4.59\\
119 & 34.4841 & -4.88158 &Y$^{4}$ & 0.54 & 1.46 & 10.0 & -3.8 & 57.2 & 2.86\\
119 (C08) &  34.4842 &  -4.88423 & & & 0.23 & -4.1 & -14.8 & 77.2 & 8.3\\
\hline
\end{tabular}
\\

{\flushleft $^{1}$ This association is designated as less reliable in the C08.\\}
{\flushleft $^{2}$ Here the best ID includes a MIPS 70$\mu$m \& 160$\mu$m association.\\}
{\flushleft $^{3}$ Here our association agrees with C08, however they note that there are multiple plausible associations, which appear to be at the same redshift so are probably associated. In each case we present data for the alternative associations below our match.\\}
{\flushleft $^{4}$ Same as $1$, but here we choose the one of the other associations.\\}
{\flushleft $^{5}$ The quoted position of this source in Table 1 of C08 is wrong.\\}

\end{table*}
As can be seen from Table \ref{tab:scubaresults} 20 of the 33 confident I07/C08 ids are reproduced exactly using our technique, 4 have different associations, but are among the multiple associations I07/C08 considered plausible\footnote{Here and throughout we define a plausible association to be one which has either a reasonable probability of being spurious ($>5$\%), or where it is not possible to discriminate between two or more possible associations}, while 9 are totally inconsistent with I07/C08.


Of the 9 discrepant associations three (SXDF850.1, SXDF850.10, SXDF850.12) are only detected in a single {\it Spitzer} band in SWIRE and hence cannot be recovered here. The alternative associations that we make for SXDF850.1 and SXDF850.12 are at separations greater 10'' and have relatively weak $\ln$ B$_{tot}$. SXDF850.10 is a nightmare scenario for the technique as it has a spurious counterpart very nearby (1.1'') with a photo-$z$ in the $1.1<z<1.4$ redshift range where 24$\mu$m is expected to be weak/non-detected due to silicate absorption.

The other six are interesting cases; SXDF850.3 has a very unusual SED with relatively bright IRAC 3.6 and 4.5$\mu$m flux but nothing in the other IRAC bands or MIPS 24$\mu$m despite the best fit template SED predicting the fluxes in these bands to be well above the detection limit. This was also noted by C08 and investigations into the origin of this unusual SED shape, whether it be real or a result of problems with the data, will be presented in future work. As a result here an alternative match at large separation is found, however this has very weak evidence ($\ln$ B$_{tot}=-5.1$).

SXDF850.29 is also a problematic association for C08. Both of the bright nearby optical sources are at low redshift ($z\approx 0.18$) and do not have SEDs suggestive of strong 850$\mu$m emission. They suggest that this could in fact be a lensed system, with the sub-mm flux coming from a background galaxy which may appear as a small amount of extended 3.6$\mu$m from one of the galaxies. Here we associate it with the nearby bright SWIRE source (2902 from C08), but with very weak evidence. While this indeed be a lensed system complicated situations such as this are unlikely to be recovered by our technique.

SXDF850.35 is a very interesting case. The SWIRE data contains detections in all 4 IRAC bands and MIPS 24$\mu$m which provides quite a good fit to the M82 template. However at the determined photo-$z$ the best fit template does not contain enough 850$\mu$m flux to be detected by SCUBA. Thus this match is given a very weak evidence ($\ln$ B$_{sed}=-103$). We make an alternative association with a slightly more distant source, but again with very weak evidence, suggesting it is not a plausible alternative. Either the C08/I07 association is wrong here, or this is a clear case of the small number of SED templates used here not being sufficient.

SXDF850.69 is designated as a less secure association by C08 as the position of the associated radio source is 13'' away from the SCUBA position. We find a weak ($\ln$ B$_{tot}=5.3$) alternative association here.

SXDF850.74 is also designated as less secure by C08. This is another relatively nearby ($z_{phot}=0.7$) optical galaxy with a very faint associated SWIRE source. Again we find a weak ($\ln$ B$_{tot}=6.1$) alternative association.

SXDF850.88 is another less secure association from C08. Here we make an association with a closer, higher $z$ object, but again with weak evidence ($\ln$ B$_{tot}=6.1$).

\begin{figure}
\centering{
\includegraphics[angle=0,scale=0.45]{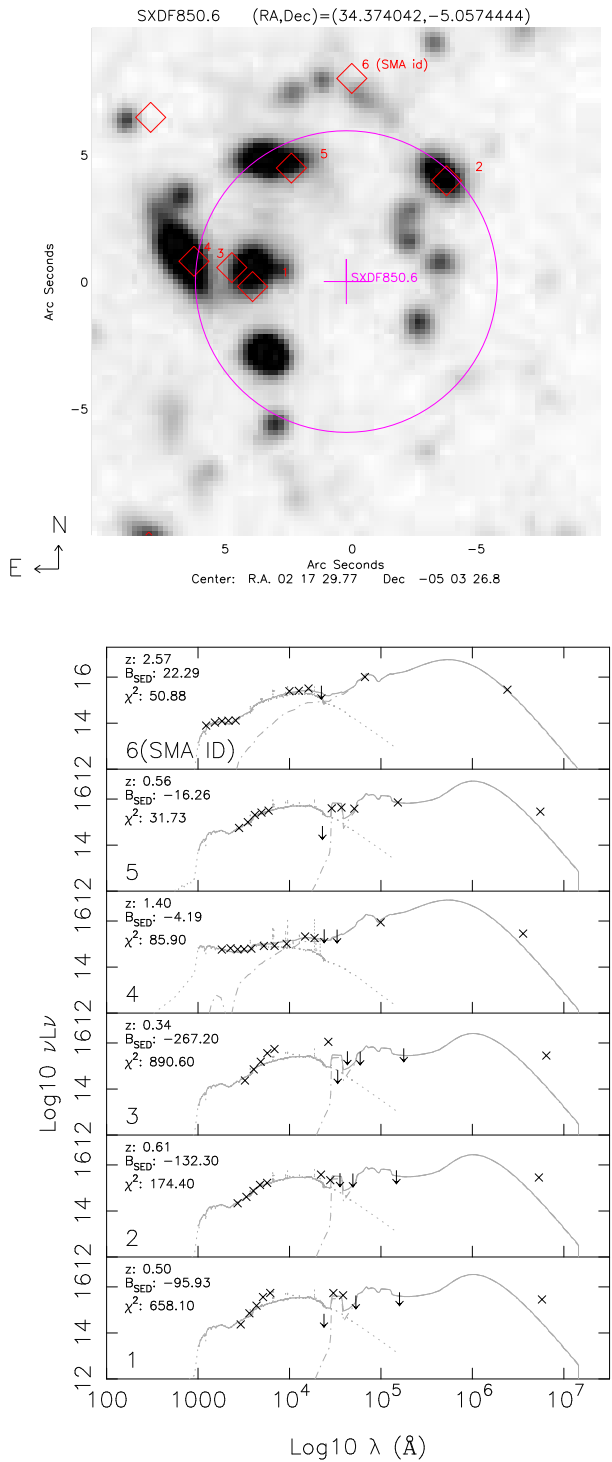}
\includegraphics[angle=270,scale=0.54]{sxdf850.6_spec.ps}

\caption{Postage stamp image of SHADES source SXDF850.6 and its surrounding area (top) and best-fit SEDs for each potential counterpart (bottom). Background is SXDS V-band image. Diamonds represent the SWIRE sources in the area. The circle represents the 6'' positional accuracy of SCUBA. For each SED; the solid line is the overall best fit optical and IR SED, the dotted line is the optical SED, while the dot-dashed line is the IR SED. The measured $\ln$ B$_{sed}$, best-fit $\chi^2$ and photo-z are also given. Where a source is undetected in the {\it Spitzer} bands the upper limit is shown. Subsequent SMA observations show that all the 850$\mu$m flux comes from ID \#6 (Iono et al. in prep).}
\label{fig:sxdf850.6}}
\end{figure}

A good example of a difficult, but correctly made, association is SXDF850.6. This source is one of the most confused scenarios and the only one in our sample to have a definitive sub-mm position from interferometric sub-mm observations with the SMA (Iono et al. in prep). Figure \ref{fig:sxdf850.6} shows the postage stamp image for SXDF850.6 and the first 6 best fit SEDs for each possible optical+SWIRE source.

Clearly SXDF850.6 is one of the most difficult cases in the sample for cross-identification. The true sub-mm source is also one of the most distant, with several other sources closer to the sub-mm position. From the values in Figure \ref{fig:sxdf850.6} it is clear that simply using the $\chi^2$ statistic would not be sufficient in this case; the lowest $\chi^2$ is given by the fourth closest which has ln B$_{sed}=-3.43$, while the true ID (\#6) has worse $\chi^2$, but significantly greater ln B$_{sed}$.
This demonstrates the power of using the Bayesian evidence, which takes into account both the likelihood of an association being the correct match, with the observed sub-mm flux {\it and} the likelihood of an association being the incorrect match, with undetected sub-mm flux.

While the results of our approach on individual sources are informative, it is worth considering the completeness and reliability statistics as presented in Section \ref{sec:scubsims}. For no cut on evidence we recover 24 of the 33 associations presented in I07/C08. For a reasonable evidence threshold, i.e. ln $B_{tot}>8$ we recover 20 I07/C08 associations with one discrepant (SXDF850.10), translating to a 64\% completeness rate, with 95\% reliability. However it is possible, if not likely, that some of the associations presented in I07/C08 are not correct. In fact C08 go so far as to indicate which associations they are not confident in; SXDF850.14, SXDF850.24, SXDF850.69, SXDF850.74 \& SXDF850.88. Of these we only recover one with reasonable evidence (SXDF850.14). If we exclude these associations from our I07/C08 ``truth'' list then our completeness improves to 72\%. Encouragingly these completeness and reliability rates are very close to those predicted from simulations in the previous section.

\begin{figure}
\includegraphics[angle=270,scale=0.4]{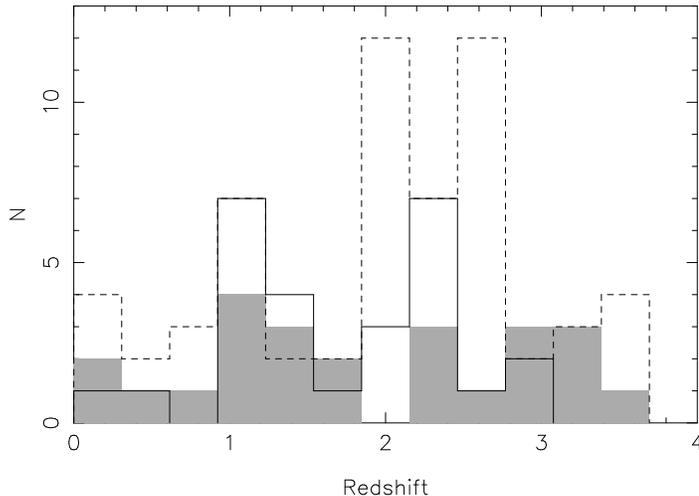}
\caption{Comparison of photometric redshift estimates for SCUBA sources from C08 (solid line), and the this work (solid grey). Also shown for comparison is the spectroscopic study of Chapman et al. 2005 (Dashed line). The median redshift of the photometric redshift samples is slightly less, but still reasonably consistent with the spectroscopic sample.}
\label{fig:zhist}
\end{figure}

A comparison of the photo-$z$ estimates between C08 and here is given in Table \ref{tab:scubaresults}. There is some level of agreement with the C08 photo-z measurements, although in a few cases the redshifts are clearly discrepant. This is more clearly seen in Figure \ref{fig:zhist}, where the distribution of both sets of photo-$z$ estimates is shown. Also shown is the redshift distribution for spectroscopically confirmed SCUBA galaxies from \nocite{Chapman2005}{Chapman} {et~al.} (2005). The median redshift for associations presented here is $z=1.73$, slightly higher than the C08 measure of the same sample ($z=1.44$) and significantly lower than the Chapman et al. sample which has a median of $z=2.5$. 

While the disagreement between the C08 and our photo-$z$ measures is troubling, this is to be expected as although the templates are similar, the photometric data and fitting algorithm are subtly different. In particular our inclusion of the UKIDSS near-IR data seems to have a significant effect on the photo-$z$ estimates. In 5 of the cases where our photometric redshift is much different than C08 (SXDF850.5, SXDF850.11, SXDF850.23, SXDF850.30 and SXDF850.71) we find that the reason for the discrepancy is that the offset from the SWIRE position and SXDS optical position is quite large ($>1.5$'') and our fit has been made using a closer UKIDSS source. Thus only the UKIDSS near-IR and {\it Spitzer} data has been used to constrain the photometric redshifts in these cases. In all but one of these cases (SXDF850.11) our photo-$z$ is much higher than the C08 estimate. Considering that these objects are possibly drop-outs in the deep SXDS $V$ band selected catalogue ($V<27.2$) high redshifts ($z>3$) should be expected.

While the median photometric redshift for 850$\mu$m sources in C08 is only slightly lower than that found here it is clear from Figure \ref{fig:zhist} that our median redshift is inflated by a small number of objects at $z>3$ and in fact the distribution is missing the large peak of galaxies at $z\sim2.5$ which appear in both the C08 and Chapman et al. analysis. There are several reasons to expect this discrepancy. As pointed out in C08 the spectroscopic redshift desert around $z\sim1.5$ slightly biases the Chapman et al. sample to higher redshifts. However the greatest effect comes from requiring a SWIRE and optical counterpart to each SCUBA source. This inherently means that higher-z sources which are too faint to be found in the SWIRE or SXDS catalogues naturally fall out of our sample. While this was less of a problem in C08, as they utilise the accompanying deep radio associations, here we rely on the quality of the SWIRE data to decide the correct association, in particular the 24$\mu$m data which does not have a large number of sources at $z>2$. Thus with the data at hand it is only possible to associate the low-z and/or high luminosity 850$\mu$m sources.


\section{Discussion}\label{sec:discussion}
\subsection{Associations for other previously unassociated SHADES SCUBA sources}
While I07 present confident radio ids for 33 SHADES SCUBA sources in the SXDF region, a total of 60 $>3.5\sigma$ 850$\mu$m sources were identified. While some of these have tenuous radio and/or mid-IR associations presented by I07, these associations are either confused or have a high chance of being spurious as determined by the p-statistic. However as discussed in Section \ref{sec:scubsims} the p-statistic can often be too harsh on sources at high redshifts where the expected counterpart would be faint.

Thus in addition to the subset of sources with confident IDs we have also run our association algorithm on the full SHADES 850$\mu$m catalogue in SXDF. Applying the same ln $B_{tot}>8$ cut used above we present 4 new plausible associations for SHADES SCUBA sources without confident radio IDs. The details of these associations are given in Table \ref{tab:newscubaresults}. 

\begin{sidewaystable*}
\label{tab:newscubaresults}
\begin{scriptsize}
\begin{tabular}{l|l|l|l|l|l|l|l|l|l|l|l|l|l|l|l|l|l|l|l|l|l}
\hline

SXDF ID  & Position & z$_{phot}$ & ln B$_{tot}$ & ln B$_{sed}$ & $\chi^2$ & F$_{3.6}$ & F$_{4.5}$ & F$_{5.8}$ & F$_{8.0}$ & F$_{24}$ & F$_{850}$ & B & V & R & $i'$ & $z'$ & J & H & K & Separation(``)\\
\hline
\\
\multicolumn{20}{|l|}{Confident}\\
\hline
32 & 34.345 -5.0126 & 2.30 & 10.69 & -0.83 & 8.36 & -- & -- & -- & 185.6 & 6050 &  27.33 & 26.81 & 26.53 & 26.20 & 25.71 & 24.31 & 24.31 & 23.47 & 7.6\\
56 & 34.446 -5.1081 & 3.67 & 42.10 &  & 268.2 &  16.97 & 16.84 & -- & -- & 166.3 & 3650 & 29.01 & 28.02 & 26.72 & 26.26 & 25.48 & -- & -- & -- & 4.9 \\ 
65 & 34.532 -5.0680 & 0.71 & 9.57 & -3.2 & 113.6 & 6.24 & 3.9 & -- & -- & -- & 4350 & 26.97 & 25.94 & 25.24 & 24.14 & 23.59 & 23.11 & 22.84 & 22.42 & 5.6\\
70 & 34.547 -5.0468 & 1.48 & 10.62 & -3.2 &23.2 & 15.94 & 14.78 & -- & -- & -- & 4050 & 25.98 & 25.79 & 25.36 & 24.84 & 24.29 & 23.01 & 22.38 & 21.92 & 2.3 \\ 
\\

\hline
\end{tabular}
\\
$^1$Association disagrees with a weak ($<40\mu$Jy) radio id from Ivison et al. (2007).\\
$^2$ Two plausible associations are found so both are quoted. \\

\end{scriptsize}

\end{sidewaystable*}


The remaining 23 SCUBA sources are again left without optical-IR associations. It is likely that the bulk of these sources are simply too faint in the {\it Spitzer} IRAC \& MIPS 24$\mu$m bands to be found in the SWIRE survey. 

\subsection{Mid-IR Properties of SHADES SMG's}
One of the distinct benefits of our association technique is that in addition to identifying the correct counterpart the best-fit photometric redshift and SED is also produced as a by-product. This allows the optical and far-IR luminosities to be easily investigated. Here we present some of these derived properties as a ``sanity'' check that our SED fitting technique is behaving as it should.

\begin{figure}
\includegraphics[angle=270,scale=0.4]{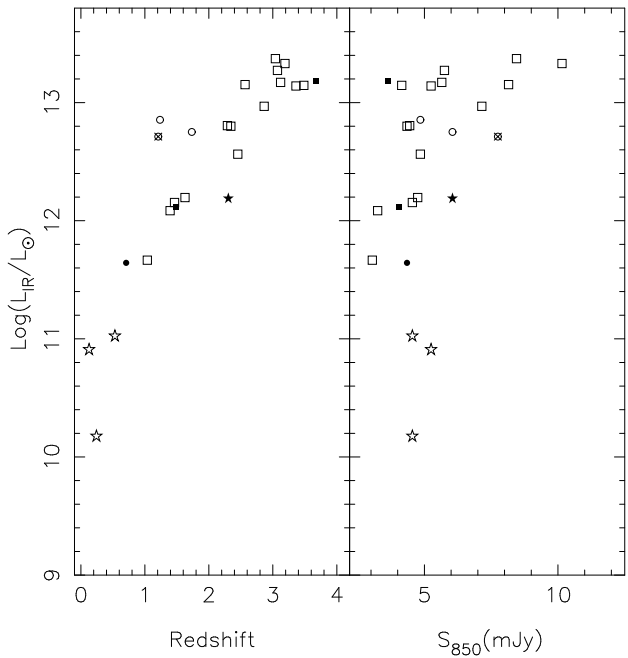}
\caption{Far-IR luminosity vs. redshift (left panel) and 850$\mu$m flux (right panel) for both the I07 SMG sample (open symbols) and the 5 new associations made here (solid symbols). SMGs are also broken down by dominate far-IR SED; Circles represent far-IR SEDs best-fit by the A220 template, Squares the M82 template, Stars the cirrus template.  }
\label{fig:lumz}
\end{figure}

Here we only include sources for which we are confident of the association via the ln $B_{tot}>8$ cut. This leaves a sample of 25 sources, the 21 from our analysis of the confident C08 sample, and the 4 new associations made in the previous section. The one known incorrect association with ln $B_{tot}>8$ (SXDF850.10) is included for the sake of fairness. Figure \ref{fig:lumz} compares the integrated far-IR luminosity (8-1100$\mu$m) to both redshift and 850$\mu$m flux. Encouragingly the far-IR luminosities measured here are predominately in the range of $10^{11}$L$_{\odot}$ to $10^{13.6}$L$_{\odot}$, consistent with both I07 and previous work on sub-mm galaxies \nocite{Ivison2002,Pope2006,Chapman2005}({Ivison} {et~al.} 2002; {Pope} {et~al.} 2006; {Chapman} {et~al.} 2005). 

Strong correlations are found between the far-IR luminosity and both redshift and 850$\mu$m flux. This is not an unprecedented result; Both Ivison et al. (2002) \nocite{Ivison2002} \& \nocite{Pope2006}{Pope} {et~al.} (2006) found similar trends in smaller samples of SMGs with photometric redshifts. Here we again conclude that the correlation with redshift of the far-IR luminosities is a result of evolution in both the number density and SED properties of ULIRG-like galaxies. 


\begin{figure}
\includegraphics[angle=270,scale=0.4]{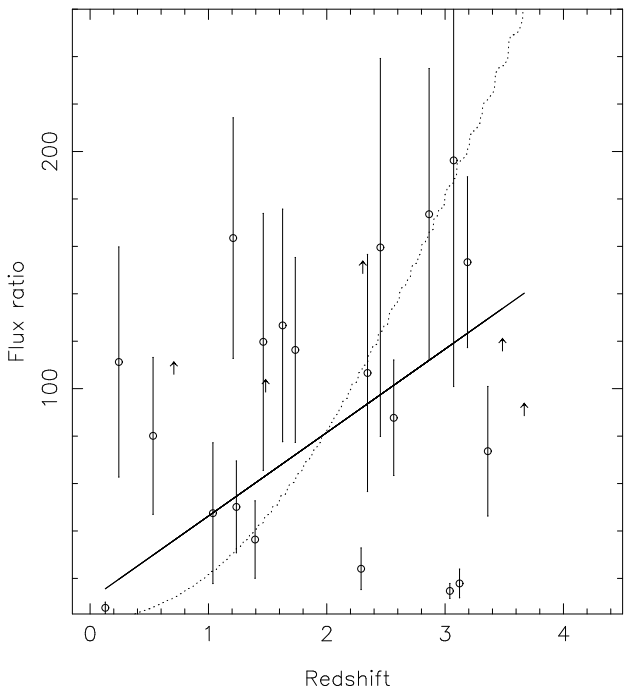}
\caption{850$\mu$m to 1.4 GHz flux ratio vs. redshift for both the I07/C08 sample and our 4 new associations. The solid line in the left panel represents the empirical relation of Chapman et al. (2005), while the dotted line represents an Arp220 template (Carilli \& Yun 2000).   }
\label{fig:irradz}
\end{figure}

Encouragingly our 4 new associations have luminosities and redshifts consistent with the rest of the SMG sample. However this does make their radio-weak nature, all 4 are undetected in deep VLA radio imaging, somewhat mysterious. To further examine whether we expect these sources to be radio-weak the 850$\mu$m-1.4 GHz flux ratio-to-redshift correlation is shown in Figure \ref{fig:irradz}. 

5 SMGs in our combined sample have no accompanying radio detection, the 4 new associations (SXDF850.32, SXDF850.56, SXDF850.65, SXDF850.70) and SXDF850.71. Of these 6, 3 have upper limits on the 850$\mu$m/1.4 GHz flux ratio which are roughly consistent with both the Chapman et al. relation and the A220 model (SXDF850.50, SXDF850.56, SXDF850.71). The other 2 all have upper limits which are higher than both predictions, i.e. they should have been detected given the depth of the radio observations. While this would appear to be a strong argument against the plausibility of these associations a large number of the I07/C08 sample of SMGs are also found to have 850$\mu$m/1.4 GHz flux ratios much higher than expected. While this may simply be a result of the large errors on both the SCUBA 850$\mu$m and 1.4 GHz radio fluxes and possibly erroneous photo-$z$'s, there are still a significant number of discrepant SMGs even when errors are taken into account. 6 SMGs from the I07/C08 sample are found to have discrepantly high 850$\mu$m/1.4 GHz flux ratios here. Of these one is the incorrect association SXDF850.10. Another three are cases where we have found a photometric redshift much less than C08 (SXDF850.8, SXDF850.11, SXDF850.37). In these cases the flux ratio would not be discrepant if the SMG is actually at the C08 photo-z estimate rather than the one made here. This leaves two cases (SXDF850.96, SXDF850.119) in which the 850$\mu$m/1.4 GHz flux ratio is inexplicably discrepant. Interestingly in these caess the p-statistic for the radio ID which is significant (0.039 \& 0.043, respectively). However both have fairly significant evidence (ln $B_{tot}=9.1$ \& 10.0 respectively. So while we hesitate to further downgrade the status of these associations, this exercise again demonstrates the diagnostic power of the radio data to discriminate between plausible associations. Additionally this also demonstrates the need for good quality redshifts, whether they be spectroscopic or, more practically, well calibrated photometric estimates.  
\subsection{Associations of SMGs in the Radio vs. the Mid-IR}
It is clear from the discussion above that deep interferometric radio images remain the most effective way to identify counterparts to sub-mm galaxies. Of the 25 IDs we are able to present with some certainty, only 5 are without radio counterparts.

\begin{figure}
\includegraphics[angle=270,scale=0.35]{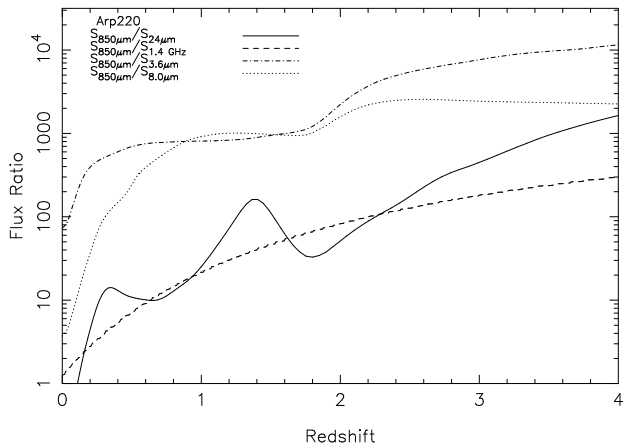}
\caption{Various flux ratios for an Arp220-like SMG at different redshifts. Expected ratios of the flux at 850$\mu$m to IRAC 3.6$\mu$m, 8.0$\mu$m and MIPS 24$\mu$m are shown as well as a model prediction for the 1.4 GHz radio flux ({Carilli} \& {Yun} 2000). }
\label{fig:a220flxrat}
\end{figure}

When considering the practicality of using mid-IR data to identify distant sub-mm sources it is worth noting the expected ratio between the sub-mm flux and those in the IRAC and MIPS 24$\mu$m bands is significantly greater than the sub-mm to radio flux ratio. This is emphasised in Figure \ref{fig:a220flxrat}. Shown are various expected flux ratios for an Arp220 template in the {\it Spitzer} IRAC and MIPS bands, and also a prediction of the 1.4 GHz radio from \nocite{Carilli2000} Carilli \& Yun (2000). It is clear that for a typical SCUBA 850$\mu$m source with S$_{850\mu m}\sim5$mJy, and an Arp220 like SED, at $z\sim2.5$ we would expect to detect it in the mid-IR at 24$\mu$m at $\sim50\mu$Jy and at 8$\mu$m at $\sim1-2\mu$Jy, while in the near-IR the 3.6$\mu$Jy flux would again be $\sim1\mu$Jy. These values are approximately an order of magnitude fainter than the nominal detection limits of the SWIRE survey. Clearly not all SMGs are so weak in the mid-IR, as the samples identified here are clearly identifiable in the SWIRE data. Deep {\it Spitzer} IRAC and MIPS surveys in fields such as GOODS and UDS do approach these depths and so we expect these datasets to be invaluable in providing counterparts for future SCUBA-2 and {\it Herschel} sources in these fields. 

Another major complication in trying to identify optical to mid-IR counterparts for sub-mm galaxies is our dependance on the template SED's to describe accurately the relationship between the flux in the different bands. In the radio this is much simplified as the far-IR -- radio correlation is known to hold to high redshifts \nocite{Chapman2005,Kovacs2006, Ibar2008}({Chapman} {et~al.} 2005; {Kov{\'a}cs} {et~al.} 2006; {Ibar} {et~al.} 2008). While here we have used a simple set of templates which are known to crudely satisfy a wide range of galaxy types (RR08), it is clear that deriving a set of SED's which properly map the properties of the sources under investigation will provide better demarcation between correct and incorrect associations. However rather than this being a failure of the techinique it could actually be its most powerful benefit as it allows for a greater flexibility in the inclusion of prior information about the galaxy population than simple techniques such as the $p$-statistic. This will be discussed further in the following section. 

\subsection{General Comments on the Applicability of Bayesian Priors based Cross-identifications}
It is clear from the work presented above that this approach is useful for situations where spatial associations are not sufficient, but reasonable priors can be assumed about the properties of the object's SEDs. In addition, as this is an automated technique which provides a single statistic (the Bayes factor) as a measure of the ``goodness'' of an association it is also very useful in situations where the number of sources requiring associations is larger than can be visually inspected. A large number of current and upcoming missions are within the bounds of these criteria, including; the BLAST experiment, SCUBA-2 legacy survey, and {\it Herschel} extra-galactic legacy surveys \nocite{Pilbratt2001}({Pilbratt} {et~al.} 2001), among many others.

As shown in Section \ref{sec:scubsims} a major failing of the $p$-statistic, and similar, techniques is that they have difficulty dealing with cases where the surface density of real counterparts is very high. This is often the case for sub-mm sources as they tend to be at high redshift and hence the short wavelength counterparts are expected to be faint, and therefore numerous. There is no way to easily modify those techniques to deal with this failing, as they are built around the notion that true association are those which could not occur by chance.

This is not a difficulty for our approach as we do not take into account the surface density of sources, but instead use our prior knowledge about SED shapes. However, the effectiveness of our technique is highly dependent on the implementation of these priors, as can be seen in the results of Section \ref{sec:scubsims} \& \ref{sec:shades}. As we wished to simply test the association algorithm the templates and priors used in the analysis are na\"{i}ve. Encouragingly even with these na\"{i}ve priors we are able to achieve similar levels of completeness and reliability as the $p$-statistic in both simulated and real data.

The real power of the technique is that it can tailored to specific applications by including priors that are more specific to the population of galaxies under consideration. One major failing of our naive set of templates is that they are not an orthogonal basis set, hence introducing a bias towards some SED shapes. For future applications a ``gold standard'' set of associations (i.e. a representative sample of associations which are known to be correct) could be used to produce a basis set of templates via Principal Component Analysis (PCA) or similar techniques. Alternatively, if a large ``gold standard'' set can be defined, the technique could be modified to do away with templates altogether and simply use the ``gold standard'' set as the model distribution to test against, in a similar way to photometric redshift technique presented by \nocite{Wolf2009} Wolf et al. (2009). In addition, all other prior information about the galaxy population in question can be included in a natural way, e.g. predicted luminosity and redshift distributions. 

It is clear that our before our approach can be used the priors on the SED must be carefully determined, tested, and optimised on mock or well known real datasets. This extra level of complexity means that existing techniques such as the $p$-statistic or other simple approaches not discussed here, such as the likelihood ratio \nocite{Sutherland1992} (Sutherland \& Saunders 1992), may be preferred for applications where the surface density of counterparts is low, and the flux distributions similar (i.e. the brightest sources match to the brightest counterparts in other catalogues). However in more difficult cases such as those present here our technique has the potential to be make reliable associations that could not be made with a simpler approach. 

%
\subsection{Applicability to Future Herschel Surveys}\label{sec:herschel}
One of the most exciting applications for our proposed technique could be finding short wavelength associations for sources detected by the instruments aboard Herschel . In particular the SPIRE instrument \nocite{Griffin2007}{Griffin} {et~al.} (2007) will be able to image at 250$\mu$m, 350$\mu$m and 500$\mu$m. However given that the primary mirror size of Herschel is 3.5m these observations will be plagued by the same issues of poor positional uncertainty as existing sub-mm facilities. In addition the BLAST experiment performed observations in the SPIRE bands utilising a smaller balloon-borne 2m primary. Thus it is of interest to see how well our technique could band merge these datasets with shorter wavelength, in particular MIPS 24$\mu$m, data.

For this exercise we construct set of mock catalogues from the GaLICS simulations which represent single band SPIRE catalogues. In addition we create a mock ``deep'' MIPS 24$\mu$m catalogue with which we want to associate our SPIRE sources. In this scenario we only consider SPIRE bands not only as these will be the most affected by positional errors, but this should also give some indications on how effective this approach would be on existing data from BLAST. The SPIRE catalogues are flux limited at S$_{250}>5$mJy, S$_{350}>7.5$mJy, S$_{500}>6.5$mJy. While these fluxes are relatively arbitrary they were chosen to match the expected source density at 250$\mu$m ($\sim 2000$ per sq. deg.) quoted by large Herschel survey programs such as HerMES (Oliver et al. in prep\footnote{\url{http://astronomy.sussex.ac.uk/~sjo/Hermes/}}). The 350$\mu$m and 500$\mu$m limits are subsequently chosen to be equivalent limits for the exposure time needed to reach the 250$\mu$m depth. For the MIPS 24$\mu$m catalogue we select all objects with S$_{24}>100\mu$Jy. These criteria result in 4 single band catalogues with 14578 24$\mu$m, 3631 250$\mu$m, 1187 350$\mu$m and 500 500$\mu$m sources respectively. Positions for sources in each single band catalogue are scattered via Gaussian random noise with a $\sigma$ equal to the expected positional uncertainty; 1.5'' for MIPS 24$\mu$m, 3.5'' for SPIRE 250$\mu$m, 4.7'' for SPIRE 350$\mu$m and 7'' for SPIRE 500$\mu$m.

While in the previous discussion the alternative hypothesis for the calculation of B$_{sed}$ was more obvious i.e. that the observed object has an SED inconsistent with the sub-mm emission, here that hypothesis is not so applicable as it is not practical to fit an SED template to the 24$\mu$m data alone. Given this, and the surprising success of the p-statistic in our previous tests, it seems natural to combine the two approaches and use the p-statistic as the alternative hypothesis. This shift is somewhat natural as the p-statistic is defined as the probability of finding a source of a given flux in a given search radius by random chance. Thus on face value it is actually the statistic we are looking for to give the probability that an association is simply a random superposition.

This introduction of the p-statistic is mildly complicated as it is designed to give the probability of finding a single source of a given flux within a given search radius, not a collection of multiple sources. To overcome this we calculate the total probability of finding our collection of sources by chance by first calculating the p-statistic for each band, given the distance from the best-estimate position, the measured flux and a search radius defined as $5\times$ the expected positional uncertainty. Then the probabilities for each band are multiplied together to give the total probability that the association is a random superposition.

We process these catalogues using our Bayesian association technique with exactly the same approach as before, except that now the alternative hypothesis in the calculation of B$_{sed}$ is the total p-statistic for the match, rather than a SED fit. The caveat that each association contains a 24$\mu$m source is introduced for practical reasons.

Processing these catalogues with the Bayesian association technique results in 3612 matches of which 3424 are 100\% correct. This translates to a completeness\& reliability of 95\%. Of the 3612 associations made 2477 are made on the basis of 250$\mu$m and 24$\mu$m data alone, while 1091 include a 350$\mu$m source, and 408 include a 500$\mu$m source. Thus while the completeness levels for the 250$\mu$m sources are very high, they drop to 91\% for the 350$\mu$m sources, and 80\% for the 500$\mu$m sources. The reliability also suffers with 94\% of 350$\mu$m associations made correctly, and only 87\% of 500$\mu$m associations made correctly. Interestingly while the error rate in the 500$\mu$m associations is disturbingly high the difference in the resulting 500$\mu$m flux quoted for the mismatch is usually very close to the value found in the correct association. In 71\% of 500$\mu$m mismatches the flux of the interloping source is within 10\% of the correct source, and only 5 (9\%) of interlopers have a flux which is more than 20\% off the true value. This is a nightmare scenario for our approach as these mismatches cannot be distinguished from the correct solution via either the SED (which would be almost the same), nor the p-statistic. However in practical terms these mismatches would be unlikely to have any discernable affect on the band merged catalogue nor scientific use of it, the 500$\mu$m sources have the worst positional uncertainty and hence contribute almost no information to the best estimate position, while the real flux uncertainty of SPIRE 500$\mu$m sources would be expected to be in the 5--10\% range. Thus while the associations of 500$\mu$m sources appears to have an unacceptably low reliability this may prove to have little consequence in terms of the scientific usage of catalogues produced in this way.

\section{Conclusion}
We have presented a new technique for finding associations between astronomical sources with large positional uncertainties. At the heart of our approach is a Bayesian framework for the association problem which extends that presented by Budav\'{a}ri \& Szalay (2008). Applications of the technique have been shown on both simulated and real sub-mm datasets from GaLICS and SCUBA, respectively.
For simulations of existing ground-based sub-mm datasets the performance of our method is found to be comparable with the p-statistic, with the key difference being that our method is superior at recovering reliable associations for the highest redshift sources.

Using a sample of SCUBA sources in SXDF from Coppin et al. (2006) with good radio identifications from Ivison et al (2007) as a testbed we recover 22 of 33 (67\%) radio identifications using only the optical to mid-IR data. Using a Bayes factor threshold it is possible to construct a catalogue with reasonable completeness (20/33:64\%) but very high reliability (95\%), successfully demonstrating the power of combining SED information and spatial information in a Bayesian way. Our technique finds plausible mid-IR associations for 4 previously unassociated SHADES SCUBA sources in the Subaru-XMM deep field. 

Finally an application of the technique on future Herschel SPIRE data is presented. We conclude that using our approach to band merge sources from the 3 SPIRE bands and {\it Spitzer} MIPS 24$\mu$m would result in merged catalogues with a completeness and reliability of $\sim 90$\%.
\section*{Acknowledgments}
Many thanks to Michael Rowan-Robinson for useful discussions regarding this work. Thanks also to Rob Ivison for useful comments which greatly enhanced the paper. \\

We thank the anonymous referee for many useful comments which greatly improved this work.\\

This work was supported by the Science and Technology Facilities  Research Council [grant number ST/F002858/1].\\

The JCMT is supported by the United Kingdoms Science and Technology Facilities Council (STFC). the National Research Council Canada (NRC), and the Netherlands Organization for Scientific Research (NWO); it is overseen by the JCMT board. We acknowledge funding support from PPARC/STFC, NRC and NASA.\\

Funding for the SDSS and SDSS-II has been provided by the Alfred P. Sloan Foundation, the Participating Institutions, the National Science Foundation, the U.S. Department of Energy, the National Aeronautics and Space Administration, the Japanese Monbukagakusho, the Max Planck Society, and the Higher Education Funding Council for England.\\

The SDSS is managed by the Astrophysical Research Consortium for the Participating Institutions. The Participating Institutions are the American Museum of Natural History, Astrophysical Institute Potsdam, University of Basel, Cambridge University, Case Western Reserve University, University of Chicago, Drexel University, Fermilab, the Institute for Advanced Study, the Japan Participation Group, Johns Hopkins University, the Joint Institute for Nuclear Astrophysics, the Kavli Institute for Particle Astrophysics and Cosmology, the Korean Scientist Group, the Chinese Academy of Sciences (LAMOST), Los Alamos National Laboratory, the Max-Planck-Institute for Astronomy (MPIA), the Max-Planck-Institute for Astrophysics (MPA), New Mexico State University, Ohio State University, University of Pittsburgh, University of Portsmouth, Princeton University, the United States Naval Observatory, and the University of Washington. \\

\bibliography{}
\appendix
\section{Summary of Budav\'{a}ri \& Szalay matching technique} \label{app:bsmath}
The association technique presented by Budav\'{a}ri \& Szalay (2008) relies on the calculation of the Bayes factor for each combination of sources from different catalogues. If we define H as the hypothesis that a set of astronomical positions from different catalogues represent the same physical source, and K the alternative hypothesis that they come from two or more sources than the Bayes factor can be written, after applying Bayes theorem as; \[B(H,K|D)=\frac{P(D|H)}{P(D|K)}\]


Budav\'{a}ri \& Szalay show that this quantity can be calculated in an iterative way over a series of catalogues via the quantities $a_k$ and $q_k$, which represent the cumulative sum of the weights, and the cumulative sum of the residuals, respectively. These quantities are calculated via the following equations,
\[a_k=a_{k-1}+w_k\]
\[q_k=q_{k-1}+\frac{a_{k-1}}{a_k}w_k\Delta_k^2\]
\[\vec{c}_k=(\vec{c}_{k-1}+\frac{w_k}{a_k}\vec{\Delta}_k)/|\vec{c}_{k-1}+\frac{w_k}{a_k}\vec{\Delta}_k|\]

\noindent Where,
\[a_k=\sum^k_{i=1}w_i\]
\[w_i=\frac{1}{\sigma^2}\]
\[\vec{\Delta}_i=\vec{x}_i-\vec{c_i}-1\]
\noindent And $\vec{c}_i$ is the unit vector of the best position for the current combination of positions,
\[\vec{c_i}=\sum^k_{i=1}w_i\vec{x_i}/|w_i\vec{x_i}|\]

Finally the logarithm of the Bayes factor, also known as the weight of evidence, is found by calculating,
\[\ln B=\ln N-\frac{1}{2}\sum^n_{i=2}q_k\]
\noindent Where,
\[N=2^{n-1}\frac{\prod{w_i}}{\sum{w_i}}\]
And the sums and products run over the $n$ catalogues.

\label{lastpage}

\end{document}